\documentstyle[twoside,fleqn,espcrc2,psfig,graphics]{article}
\newcommand{\AmS}{{\protect\the\textfont2
  A\kern-.1667em\lower.5ex\hbox{M}\kern-.125emS}}

\title{The Maximal Abelian Gauge in SU(3) Lattice Gauge Theory
}

\input epsf

\author{William W. Tucker and John D. Stack
\address{Department of Physics, University of Illinois at 
Urbana-Champaign, 1110 West Green Street, Urbana, Illinois, 68101}
\thanks{This work was supported in part by the National Science 
Foundation.  The computations on the SV1 system (Texas, Austin)
were supported by NPACI.
}}
       
\begin{document}

\begin{abstract}
We gauge fix 600 $SU(3)$ $\beta=6.0$ configurations on a $16^4$ lattice 
to a simple form of the maximal abelian gauge.
We project the $SU(3)$ valued links
to the $U(1)\times U(1)$ subgroup, and extract $U(1) \times U(1)$ and
monopole string tensions.  After gauge fixing to the indirect center gauge,
the $U(1) \times U(1)$ links are projected to $Z(3)$ and a vortex
string tension is measured.  The vortex and magnetic current densities are
measured.
\end{abstract}

\maketitle

The idea that the long-range physics of an $SU(N)$ gauge
field can be compressed into the abelian sector of that field
was suggested by {\mbox 't Hooft} \cite{thooft}.
The simplest way to do this is to gauge fix the field so that
it is as much in the abelian ``direction" as possible and then to
project to the abelian sector.

The process of gauge fixing to the maximum abelian gauge (MAG) 
has been studied extensively in $SU(2)$ lattice gauge theory.
Gauge fixed configurations are projected to the $U(1)$ subgroup.
From this $U(1)$ gauge field, monopole currents are located and
a potential from the $U(1)$ gauge field or from the monopoles can
be found and compared to the full $SU(2)$ case.
Also studied is the indirect center gauge  (ICG). Here, after going to the
MAG, the
$U(1)$ field is also gauge fixed and then projected to the $Z(2)$ center
of the group \cite{greensite2}.
There are still unresolved issues such as stability under cooling
and sensitivity to gauge ambiguities for the case of $SU(2)$ \cite{jsosaka}.
It is natural, however, to go on to the case of an $SU(3)$
gauge group in order to look at confinement as it occurs in QCD.
\vspace{-0.1in}
\section{ Maximal Abelian Gauge}

In $SU(2)$ gauge theory in the continuum, 
the MAG is the gauge that minimizes the quantity
\begin{equation}
G_{mag\_c}^{SU(2)}
=\int{\left[(A_\mu^1)^2+(A_\mu^2)^2\right]d^4x}.
\label{contmag2}
\end{equation}
The equivalent lattice functional is

\begin{equation} 
G_{mag\_l}^{SU(2)}=\frac{1}{2N_{link}}\sum_{x,\mu}
{\rm Tr}\left[U^{\dagger}_{\mu}(x)\sigma_{3}U_{\mu}(x)
\sigma_{3}\right].
\label{latmag2}
\end{equation}
A maximum of Eq.(\ref{latmag2})  is reached by sweeping over 
the lattice and performing gauge
transformations that maximize the lattice functional at a given site.

In $SU(3)$, the corresponding continuum functional is
\begin{eqnarray}
G_{mag\_c}^{SU(3)} & = & \int{\left[(A_\mu^1)^2+(A_\mu^2)^2+(A_\mu^4)^2+(A_\mu^5)^2 + \right.} \nonumber\\
& & \left. (A_\mu^6)^2+(A_\mu^7)^2\right]d^4x.
\label{contmag3}
\end{eqnarray}
On the lattice Eq.(\ref{contmag3}) becomes
\begin{eqnarray}
G_{mag\_l}^{SU(3)}& = & \frac{1}{4N_{link}}\left\{
\sum_{x,\mu}
{{\rm Tr}\left[U^{\dagger}_{\mu}(x)\lambda_{3}U_{\mu}(x)
\lambda_{3}\right]}+\right. \nonumber \\
& &\left.\sum_{x,\mu}{{\rm Tr}\left[U^{\dagger}_{\mu}(x)\lambda_{8}U_{\mu}(x)
\lambda_{8}\right]} \right\}.
\label{latmag3}
\end{eqnarray}
For $SU(3)$, gauge fixing updates are done based on $SU(2)$ subgroups.
We denote the three subgroups as follows:
$SU(2):\,\lambda_{1},\lambda_{2},\lambda_{3},$ 
$SU(2)^{\prime}:\,  \lambda_{4},\lambda_{5},\lambda^{\prime}_{3},$
$SU(2)^{\prime\prime}:\, \lambda_{6},\lambda_{7},\lambda^{\prime\prime}_{3}$,
where $\lambda_{i}$ are the usual $\lambda$-matrices, and 
$\lambda_3^\prime={\rm diag}(1,0,-1)$ and
$\lambda_3^{\prime\prime}={\rm diag}(0,1,-1)$.

\section{Abelian Projection}

Having reached a maximum of Eq.(\ref{latmag3}), it is necessary to project
each  $SU(3)$ link matrix to the $U(1)\times U(1)$ subgroup.
In the $SU(2)$ case, the projection to $U(1)$ is determined by the (opposite)
phases of the diagonal elements of each link matrix.
For $SU(3)$, the phases of the diagonal elements of the link matrices do not
determine a diagonal unitary $3\times 3$ matrix.
One simple way to project to $U(1) \times U(1)$ is to choose two of the
diagonal phases, and determine the third by unitarity.
Alternatively, each diagonal phase can be used, with unitarity enforced
by an additional overall phase.
The method used here is to find a diagonal matrix $U$ such that the quantity
${\rm Re}\left({\rm Tr}\left[UA^{\dag} \right] \right)$
is maximized, where $A$ is the $SU(3)$ link.
The purpose of this method is to find the ``best" $U(1)\times U(1)$
approximation to a given $SU(3)$ matrix.
Either of the two methods just described can be used for a trial solution.
The trial solution is then successively  updated using matrices of the
form $\exp(i\phi\lambda_{3})$, or matrices with   $\lambda_{3}^{\prime},
\lambda_{3}^{\prime\prime}$ replacing $\lambda_{3}$.
The procedure was found to give numerical accuracy in
$10$ iterations.
As a side note, it  would be the natural procedure to use for projecting
spatially smeared $U(1)\times U(1)$ links back to the $U(1)\times U(1)$
manifold.

\section{Indirect Center Gauge}

With configurations projected to the $U(1)\times U(1)$ subgroup,
it becomes a simple (and considerably less computationally intensive)
matter to gauge fix to the indirect center gauge (ICG).
This gauge is determined by maximizing the functional

\begin{equation}
G_{icg}^{SU(3)}=\frac{1}{9N_{link}}\sum_{x,\mu}
\left|{\rm Tr}\left[U_{\mu}(x)\right]\right|^{2},
\label{icgfun2}
\end{equation}
where the $U_{\mu}$ are the $U(1)\times U(1)$ projected links.
The easiest way to reach the ICG is with gauge transformations by factors
of $\tilde{\lambda}_8$, $\tilde{\lambda}_8^\prime$, and
$\tilde{\lambda}_8^{\prime\prime},\,$ where
$\tilde{\lambda}_8=\sqrt{3}\lambda_8={\rm diag}(1,1,-2)$,
$\tilde{\lambda}_8^\prime={\rm diag}(1,-2,1)$, and
$\tilde{\lambda}_8^{\prime\prime}={\rm diag}(-2,1,1)$.
This choice of transformation has the advantage that solving for
the optimal transformation gives an easy analytical solution.

Projection to $Z(3)$ is accomplished by choosing the element
$\xi^j{\bf 1}, j=0,1,2,\, \xi=\exp(2\pi i/3)$ that maximizes the quantity
${\rm Tr}[\xi^j{\bf 1}U_{\mu}^\dagger]=\xi^j{\rm Tr}[U_{\mu}^\dagger]$, for
each $U(1) \times U(1)$ link.
This results in a set of $Z(3)$ links. The Wilson loop is then calculated
by multiplying these $Z(3)$ valued links around the loop.

\section{Magnetic Currents}

Magnetic currents are extracted for each $SU(3)$ color by applying the
Toussaint DeGrand procedure to the $U(1) \times U(1)$ links.  This 
produces three magnetic currents, of which only two are 
independent.  In our
method \cite{jrsu3}, the sum over colors of the magnetic current 
is non-vanishing on a link
by link basis.  Nevertheless, when used to calculate a Wilson loop, this total
current produces a null potential, as it should.  After extracting the 
magnetic currents, monopole Wilson loops are calculated for each color.  This
part of the calculation proceeds exactly as in $U(1)$ or $SU(2)$ lattice
gauge theory \cite{js-rw,stack}.  
Finally the color average of these monopole Wilson loops 
is used to produce a monopole potential from which a monopole string tension
is extracted.    The color averaged fraction of links carrying magnetic current 
was also measured and found to be $7.44(4) \times 10^{-3}$.

\begin{table*}[htb]
% space before first and after last column: 1.5pc
% space between columns: 3.0pc (twice the above)
\setlength{\tabcolsep}{3.5pc}
% -----------------------------------------------------
% adapted from TeX book, p. 241
\catcode`?=\active \def?{\kern\digitwidth}
% -----------------------------------------------------
%\caption{String tensions for SU(3) MAG}
\caption{Potential $V(R)=\sigma R + \alpha/R + V_0$}
\label{tab1}
\begin{tabular*}{\textwidth}{@{}l@{\extracolsep{\fill}}rccc}
\hline
%                 & \multicolumn{2}{l}{Pilot plant} 
%                 & \multicolumn{2}{l}{Full scale plant} \\
%\cline{2-3} \cline{4-5}
                 & \multicolumn{1}{r}{$Potential$}
                 & \multicolumn{1}{c}{$\sigma$}
                 & \multicolumn{1}{c}{$\alpha$}
                 & \multicolumn{1}{c}{$V_0$}  \\
\hline
& $SU(3)$ & 0.053 &   -0.267    &  0.625 \\
& $U(1)\times U(1)$ & 0.045(2) &   -0.07(2)    & 0.19(1) \\
& ICG & 0.040(2) &   0.03(1)    &  -0.02(1)\\
& Monopole & 0.037(1) &   0.02(1)    &  -0.04(1)\\

\hline
%\multicolumn{5}{@{}p{120mm}}{Reprinted from: G.M. Ritcey,
%                             Tailings Management,
%                             Elsevier, Amsterdam, 1989, p. 635.}
\end{tabular*}
\end{table*}

% \vspace{-0.3cm}

\section{Results}

We gauge fixed each of 600 configurations to the MAG
using Eq.(\ref{latmag3}), generating
one gauge copy per configuration.
The MAG gauge fixing was done using overrelaxation \cite{mandula} with
overrelaxation parameter $\omega=1.8$.
Gauge fixing was continued until $\left<| X_{12}|^2+|
X_{13}^{\prime}|^2+| X_{23}^{\prime\prime}|^2\right> < 3\times10^{-11}$
was satisfied where the matrix $X(x)$ is 
\begin{equation}
\sum_{\mu}U_{\mu}(x)\lambda_{3}U_{\mu}^{\dagger}(x)+
 U_{\mu}^{\dagger}(x-\mu)\lambda_{3}U_{\mu}(x-\mu),
\label{magstop}
\end{equation}
and the matrices $X^{\prime},X^{\prime\prime}$ are similar, but 
have $\lambda_{3}^{\prime},
\lambda_{3}^{\prime\prime}$, respectively in place of $\lambda_{3}$.

Further gauge-fixing using Eq.(\ref{icgfun2}) was performed to get
to the ICG.
The stopping criterion for this procedure was the same as in \cite{jrsu3},
One gauge copy/configuration was generated, and the overrelaxation parameter
for ICG was $\omega=1.7$.
Links were then projected to $Z(3)$ and Wilson loops calculated
as in a $Z(3)$ gauge theory.
The P-vortex density, that is the fraction of  $Z(3)$ plaquettes with non-zero
flux, was measured to be $2.25(1) \times 10^{-2}$. 

\begin{figure}[htb]
\psfig{file=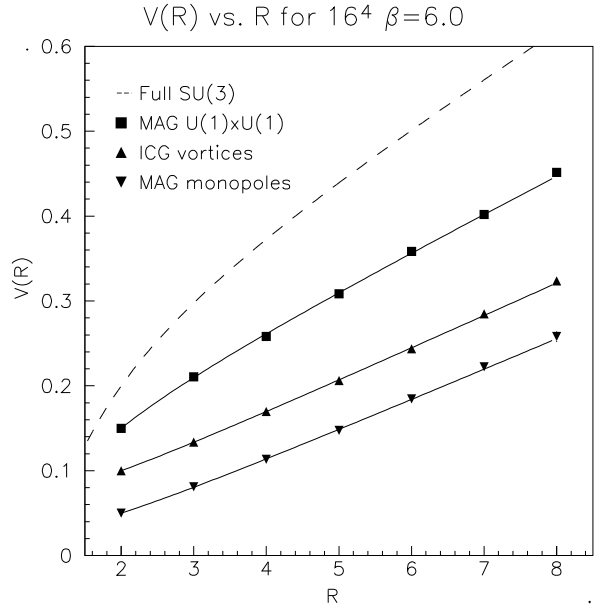,height=8.cm}
\vspace{-1.0cm}
\caption{V(R) vs R for MAG ($U(1) \times U(1)$ and monopoles)
and ICG \mbox{(P-vortices)} for $SU(3)$}
\label{fig1}
\end{figure}

Wilson loops up to $R=8,T=8$ were generated for the various cases.
Potentials were determined from Wilson loop values by
fitting $-\ln(W(R,T))$ to a straight line in $T$. 
The $\ln(W)$  vs $T$ plots
were remarkably linear in all cases, even for $T < R$. From the potentials,
string tensions were
determined by fitting to the form $V(R)=\sigma R+\alpha/R+V_0$.
Values for these parameters are given in Table \ref{tab1}.
As can be seen there, the MAG $U(1) \times U(1)$, ICG vortex, and MAG monopole
string tensions all lie below the full $SU(3)$ value \cite{bali} 
by an amount well
outside of error bars.  In Fig.(\ref{fig1}), we plot the potentials determined
here along with the best  $SU(3)$ potential from the literature \cite{bali}.   
The values of the constant $V_{0}$ have been adjusted for display.

Our findings for $SU(3)$ are  to be contrasted with the behavior in
$SU(2)$, where for one gauge copy/configuration, 
at  couplings with comparable string tension, the
MAG $U(1)$ and $ICG$ vortex string tensions are larger than the full
$SU(2)$ value \cite{jsosaka}, 
while the monopole string tension lies very close to
the full $SU(2)$ value \cite{stack}.  Accounting for gauge copies with higher
values of the functionals causes all
the string tensions to decrease, with the $U(1)$ MAG string tension ending up
$\sim 8\%$ below the full $SU(2)$ result \cite{bali1}.  
If the same trend holds here,
the $U(1) \times U(1)$ MAG string tension would end up well
below the full $SU(3)$ value, with the monopole and ICG vortex 
values lower still.  To remedy this situation for $SU(3)$,
we are studying more general forms of the MAG condition \cite{jrsu3}.

\end{document}